\documentclass[%
pre, amsmath,amssymb,showkeys,
aps,superscriptaddress,twocolumn,a4paper,english]{revtex4-2}
\usepackage[T1]{fontenc}
\usepackage[utf8]{inputenc}
\usepackage{amsmath}
\usepackage{amssymb}
\usepackage{graphicx}
\usepackage{bm}
\usepackage{refstyle}
\usepackage{multirow,array}
\usepackage{mathtools}
\usepackage{empheq}
\usepackage{chessfss}
\usepackage{epsdice}
\usepackage{pifont}
\usepackage{amsthm}
\usepackage[
citecolor=blue,
colorlinks,
linkcolor=blue,
urlcolor=blue,
]{hyperref}
\usepackage[dvipsnames]{xcolor}
\usepackage{dcolumn}
\usepackage{bm}

\usepackage{soul} 
\usepackage{cancel}
\begin{document}

\title{Oscillatory evolutionarily stable state and limit cycle in replicator dynamics}

\author{Suman Chakraborty}
\email{ph25r001@smail.iitm.ac.in}
\affiliation{
	Department of Physics, Indian Institute of Technology Madras, Chennai 600036, India
}
\author{Vikash Kumar Dubey}
\email{vdubey9818@gmail.com (corresponding author)}
\affiliation{
	Department of Physics, Indian Institute of Technology Kanpur, Uttar Pradesh 208016, India
}
\author{Vaibhav Madhok}
\email{ madhok@physics.iitm.ac.in}
\affiliation{
	Department of Physics, Indian Institute of Technology Madras, Chennai 600036, India
}

\author{Sagar Chakraborty}
\email{sagarc@iitk.ac.in}
\affiliation{
	Department of Physics, Indian Institute of Technology Kanpur, Uttar Pradesh 208016, India
}

\date{\today}
\begin{abstract}
The idea of evolutionarily stable state (ESS) of a population is a cornerstone of evolutionary game theory; moreover, it coincides with the game-theoretic concept of Nash equilibrium. Such a state corresponds to a strategy adopted by the population such that a rare mutant strategy cannot invade the population. In parallel, the dynamical formulation of evolutionary game theory---particularly through replicator dynamics embodying the tenet of survival of the fittest---provides a framework for modelling frequency-dependent selection over time. While it is well known that an ESS corresponds to stable fixed point in replicator dynamics, the evolutionary game-theoretic characterization of limit cycles is unknown. Here we fill this lacuna by defining oscillatory ESS (OESS) which we prove to be a stable limit cycle. We also show when an OESS is unique and if there are multiple OESSes, then what their locations are in the phase space.  
\end{abstract}
\maketitle
\section{Introduction}
Evolutionarily stable strategy~\cite{SMITH1973nature}---a strategy adopted by a resident population that no other rare mutant strategy can invade it---is the most fundamental concept of evolutionary game theory~\cite{smith_1982, Thomas1985jmb}, and has been hailed~\cite{dawkins1989selfish} as second only to the Darwin’s contribution in advancing the evolutionary theory. It manifests as evolutionarily stable \emph{state} (ESS) that specifies robust fractions of types (marked by strategies) of individuals in the population: ESS is locally asymptotically stable fixed point of replicator equation~\cite{Taylor1978mb, Schuster1983jtb} that models the evolution of population state under Darwinian tenet of natural selection (survival of the fittest~\cite{Spencer1864,Darwin1869}), i.e., the per capita growth rate of a type is positive if its fitness is more than that of the population's average fitness. Thus, with ESS one can associate both game-theoretical and dynamical interpretations.

However, more complex evolution of states---rather than mere convergence onto some fixed state (ESS)---is not unheard of: Some famous examples are oscillations in the frequencies of two cichlid fish’s phenotypes~\cite{Hori1993science}, oscillations
in the frequencies of different throat-colour phenotypes of side-blotched lizards~\cite{Sinervo1996nature, Sinervo2000nature} and oscillations in population of three different strains of \emph{E.~coli} bacteria~\cite{Kirkup2004}. Mathematically, a dynamically stable oscillatory state is an attractor that is not just a point and so it is not clear how to associate a game-theoretical meaning to it \emph{\`a la} ESS. In passing, a related interesting fact is worth mentioning: Many ecological systems---modelable by the
corresponding Lotka-Volterra equation~\cite{Lotka1920pnas}---are known to exhibit cyclic
evolution of the abundance of the constituent species~\cite{Peterson1984science,Fauteux2015jae}, and the Lotka-Volterra equation is mathematically mappable~\cite{hofbauer_book} onto the replicator equation~\cite{Taylor1978mb, Schuster1983jtb} which is our focus in this paper.

While some of the recent works~\cite{Mukhopadhyay2020jtb,Mukhopadhyay2020chaos,Bhattacharjee2023pre,Dubey2024dg} have tried to provide a game-theoretical interpretation of oscillatory outcomes in the context of the discrete-time version of the replicator equation, the question is open for the case of continuous-time version of the replicator equation. While with only two strategies in a population, oscillatory solutions are possible in the former; the latter requires at least four strategies to show limit cycle outcome~\cite{Hofbauer1981}. This is quite a technical challenge; moreover,  exact analytical form of a limit cycle solution is unknown in most cases. Nevertheless, as we shall see below, this paper succeeds in not only proposing a compelling extension---termed  Oscillatory Evolutionarily Stable State (OESS)---of ESS to limit cycles but in also making sure that all the major desired connections between game-theoretical and dynamical interpretations of ESS are smoothly extended to OESS.

\section{OESS}
Let us start by recalling the replicator equation for a population with $n$ types (strategies):
\begin{eqnarray}\label{eq:replicator}
	\frac{dx_i}{dt} = x_i(t) \left[({\sf A}\boldsymbol{x}(t))_i - \boldsymbol{x}(t)\cdot {\sf A}\boldsymbol{x}(t) \right]~ \forall i \in \{1, \cdots, n\},\quad\,\,\,
\end{eqnarray}
where ${\sf A}$ is the payoff matrix representing the population game, $x_i(t)$ is the fraction of the $i$th type at time $t$, $\boldsymbol{x}(t) \in S_n \equiv \left\{ (x_1, \cdots, x_n) \in \mathbb{R}^n \mid x_i \geq 0, \sum_{i=1}^{n} x_i = 1 \right\}$ is the population state in the $(n-1)$-dimensional simplex, and $\boldsymbol{x}(t)\cdot{\sf A}\boldsymbol{x}(t) = \sum_{j=1}^{n} x_j(t) ({\sf A}\boldsymbol{x}(t))_j$ is the average fitness of the population.

Let us consider that the number of strategies are more than three ($n>3$) so that there may exist a limit cycle~\cite{Hofbauer1981,hofbauer_book}---an isolated, closed trajectory in the phase space---with time-period $\tau>0$. Notationally, a phase point on  the limit cycle satisfies
$\hat{\boldsymbol{x}}(t+\tau) = \hat{\boldsymbol{x}}(t)$.
Since the replicator dynamics renders the $(n-1)$-dimensional simplex and its constituent $(n-2)$-dimensional subsimplices forward invariant, the support of a closed trajectory cannot change with time; thus, without any confusion, we may denote that support as $\text{supp}(\hat{\boldsymbol{x}})$ without any explicit mention of time. We note that since the vertices of the simplex phase space are fixed points of the replicator dynamics, they cannot be a phase point on any limit cycle. 

For notational convenience, any piece of phase trajectory from time $t_0$ to $t_1$ will be represented as a continuous sequence of states: $\{\boldsymbol{x}(t)\}_{t\in [t_0,\, t_1]}$. Therefore, a complete periodic trajectory would read $\{\hat{\boldsymbol{x}}(t)\}_{t\in [t_0-\tau,\, t_0]}$.
However, due to periodic nature, the choice of the initial time \(t_0\) is practically irrelevant for our purposes: the sequence is collection of the same set of values for any \(t_0\); therefore, 
for a periodic trajectory, we may further simplify the notation and write the sequence compactly as $\{\hat{\boldsymbol{x}}(t)\}_{\tau}$.

Recall that in the classical setting, a particular state \(\hat{\boldsymbol{x}}\) is called an
ESS if $\hat{\boldsymbol{x}}\cdot {\sf A}\,\boldsymbol{x} 
- \boldsymbol{x}\cdot{\sf A}\,\boldsymbol{x} > 0$
for all \(\boldsymbol{x}\) in some deleted neighbourhood of \(\hat{\boldsymbol{x}}\); and this happens to corresponds to local attraction under the replicator dynamics. So now we ask how to formulate a reasonable concept of evolutionary stability for limit cycles which (i) reduces to the known concept of ESS as $\tau\to0$ and (ii) is dynamically stable as well. We take hint from analogous studies~\cite{Mukhopadhyay2020jtb,Mukhopadhyay2020chaos,Bhattacharjee2023pre,Dubey2024dg} on discrete-time replicator equation and guess that the effective fitness of a state on limit cycle should also depend on other states on it on average. The validity and justification of the definition would appear more reasonable {\it a posteriori} in the light of the Theorem presented later in this paper.

For this purpose, consider the average fitness of the resident when it is in the state  $\hat{\boldsymbol{x}}(t_0)$ at time $t_0$ to depend on how the resident state has evolved over one complete cycle  $[t_0-\tau, t_0]$. This average fitness is given by the following time average,
\begin{eqnarray}\label{eqn:time_average_payoff}
 \frac{1}{\tau} \int_{t_0-\tau}^{t_0} \hat{\boldsymbol{x}}(t_0)\cdot{\sf A}\hat{\boldsymbol{x}}(t)\, dt
 =\hat{\boldsymbol{x}}(t_0)\cdot{\sf A}	\hat{\bar{\boldsymbol{x}}},
\end{eqnarray}
where bar on the top of a symbol represents time-average over time-period $\tau$. For example, more explicitly,
\begin{equation}
	\hat{\bar{\boldsymbol{x}}}\equiv\frac{1}{\tau} \int_{t_0-\tau}^{t_0} \hat{\boldsymbol{x}}(t)\, dt.
\end{equation}
We note that $\hat{\bar{\boldsymbol{x}}}$ depends on $\tau$ (which we shall keep implicit through the use of overbar throughout the paper) but not on $t_0$. Finally, averaging Eq.~(\ref{eqn:time_average_payoff}) over one full period gives the time–average fitness of a resident \emph{evolving} on the limit cycle: 
\begin{eqnarray}
	\frac{1}{\tau} \int_{0}^{\tau} \hat{\boldsymbol{x}}(t_0)\cdot{\sf A}\hat{\bar{\boldsymbol{x}}}\, dt_0
	=\hat{\bar{\boldsymbol{x}}}\cdot{\sf A}	\hat{\bar{\boldsymbol{x}}}.
\end{eqnarray}
One may aptly term this as fitness of the limit cycle where double-averging---one over the resident as focal state and the other over the opponent state---has been performed; after all, both are periodically evolving with time on the limit cycle. We extend this idea of the limit cycle's fitness to the case when the opponent state is that of a mutant, denoted by $\boldsymbol{x}$:
\begin{eqnarray}\label{eqn:fitness_res_mut}
	\frac{1}{\tau} \int_{0}^{\tau}\left[\frac{1}{\tau}\int_{t_0-\tau}^{t_0}\hat{\boldsymbol{x}}(t_0)\cdot{\sf A}\boldsymbol{x}(t)dt\right] dt_0 \qquad\qquad\qquad\qquad\nonumber\\
	=\frac{1}{\tau}\int_{0}^{\tau}\hat{\boldsymbol{x}}(t_0)\cdot{\sf A}	\bar{\boldsymbol{x}}(t_0)dt_0.\quad
\end{eqnarray}
Note that we may not remove the dependence on $t_0$ from $\bar{\boldsymbol{x}}(t_0)$ in the last equality, unlike in Eq.~(\ref{eqn:time_average_payoff}), because $\{\boldsymbol{x}(t)\}$ may not be periodic.
 
Equipped with the idea of the fitness of limit cycle, we can now introduce the notion of evolutionary stability. To do so, we assume that the mutant remains rare, occupying a fraction $\epsilon$, always below an invasion barrier $\bar{\epsilon}(\boldsymbol{x}(t))$; i.e., at any time $t$, the mutant fraction satisfies $\epsilon < \bar{\epsilon}(\boldsymbol{x}(t))$. In this setting, the following definition specifies the criterion for evolutionary stability, which we refer to as the oscillatory ESS of period $\tau$ (\text{OESS}$_\tau$):
\\
\\
\noindent\textbf{Definition 1a}: A sequence  of states  $\{\hat{\boldsymbol{x}}(t)\}_\tau$ is called OESS$_\tau$ if for any mutant state sequence $\{\boldsymbol{x}(t)\}_{t \in [-\tau,\tau]}$, different from the resident, there exists an invasion barrier $\bar{\epsilon}(\boldsymbol{x}(t))>0$ such that $\forall \epsilon$  with $0 < \epsilon <\bar{\epsilon}(\boldsymbol{x}(t))$,
\begin{eqnarray}\label{eq:ess_general}
	\frac{1}{\tau} \int_{0}^{\tau} \hat{\boldsymbol{x}}(t_0)\cdot {\sf A} \big[(1 - \epsilon) \hat{\bar{\boldsymbol{x}}}+ \epsilon \bar{\boldsymbol{x}}(t_0)\big]~dt_0
	>\nonumber\\
	\frac{1}{\tau} \int_{0}^{\tau} \boldsymbol{x}(t_0)\cdot {\sf A}   \left[ (1 - \epsilon) \hat{\bar{\boldsymbol{x}}}  +\epsilon \bar{\boldsymbol{x}}(t_0)\right]~dt_0.
\end{eqnarray}
We refer to this as the first-principle definition of OESS. Effectively, it is saying that in a population where periodically evolving residents  are mixed with some mutants who also temporally evolve, the time-averaged fitness of a resident against this mix is more than that of a mutant. Following two alternative versions (see Appendix~\ref{sec:equivalence_state} for the proof of equivalence) of this definition turn out to be more useful in subsequent sections for mathematically proving various results.
\\
\\	
\noindent\textbf{Definition 1b}: A sequence  of states $\{\boldsymbol{\hat{x}}(t)\}_\tau$ is an OESS$_\tau$ if for any mutant state sequence $\{{\boldsymbol{x}}(t)\}_{t \in [-\tau,\tau]}$, the following conditions hold true:
\begin{equation}\label{eqn:equlibrium_1b}
	\text{(i)}~\frac{1}{\tau} \int_{0}^{\tau}
	\hat{\boldsymbol{x}}(t_0)\cdot{\sf A}\hat{\bar{\boldsymbol{x}}}~dt_0
	\ge \frac{1}{\tau} \int_{0}^{\tau}
	\boldsymbol{x}(t_0)\cdot  {\sf A} \hat{\bar{\boldsymbol{x}}}~dt_0
\end{equation}
(ii) and for those $\{{\boldsymbol{x}}(t)\}_{t \in [-\tau,\tau]}\neq\{\boldsymbol{\hat{x}}(t)\}_\tau$ for which inequality in (i)  holds as equality,
\begin{equation}\label{eq:ess_stab}
	\frac{1}{\tau} \int_{0}^{\tau} \hat{\boldsymbol{x}}(t_0)\cdot  {\sf A} \bar{\boldsymbol{x}}(t_0)~dt_0
	>
	\frac{1}{\tau} \int_{0}^{\tau} \boldsymbol{x}(t_0)\cdot {\sf A}\bar{\boldsymbol{x}}(t_0)~dt_0.
\end{equation}
\noindent\textbf{Definition 1c}: A state sequence  $\{\hat{\boldsymbol{x}}(t)\}_\tau$ is an OESS$_\tau$ if for all the sequence of states $\{\boldsymbol{x}(t)\}_{t \in [-\tau,\tau]}$ in some deleted neighborhood of $\{\hat{\boldsymbol{{x}}}(t)\}_\tau$, the following holds true:
\begin{equation}\label{eq:OESS_neighbourhood}
	\frac{1}{\tau} \int_{0}^{\tau} \hat{\boldsymbol{x}}(t_0)\cdot  {\sf A} \bar{\boldsymbol{x}}(t_0)~dt_0
	>
	\frac{1}{\tau} \int_{0}^{\tau} \boldsymbol{x}(t_0)\cdot {\sf A} \bar{\boldsymbol{x}}(t_0)~dt_0.
\end{equation}

Finally, we can satisfactorily see that when the sequence has the constant value at all times, which trivially implies a fixed point, all the above versions of the OESS definition reduce to the corresponding ones known for ESS~\cite{hofbauer_book}. Also, just like in the case of ESS, a natural question comes up about the number of OESSes a system can have. We now present two propositions (Proposition~1a and Proposition~1b) that help us to answers such question.
\section{Existence of OESS}
\noindent\textit{\textbf{Proposition~1a:} Suppose there exists an OESS $\{\hat{\boldsymbol{x}}(t)\}_\tau$ and consider another state sequence $\{\hat{\boldsymbol{x}}'(t)\}_{\tau'}$ that cannot be obtained from $\{\hat{\boldsymbol{x}}(t)\}_\tau$ by a time-scaling transformation. If  $\text{supp}(\hat{\boldsymbol{x}}')\subseteq\text{supp}(\hat{\boldsymbol{x}})$, then  $\{\hat{\boldsymbol{x}}'(t)\}_{\tau'}$ is not an OESS.}\\
\emph{Proof.} Let us start by assuming that $\{\hat{\boldsymbol{x}}'(t)\}_{\tau'}$ is a potential OESS such that $\text{supp}(\hat{\boldsymbol{x}}')\subseteq\text{supp}(\hat{\boldsymbol{x}})$, i.e., if there exists an  index $i$ such that $i \in \text{supp}(\hat{\boldsymbol{x}}')$, then 
$i \in \text{supp}(\hat{\boldsymbol{x}})$.

Let us define $s_0 = \frac{\tau' t_0}{\tau}$. In the light of this, we introduce a notation $\hat{\boldsymbol{y}}$ such that $\hat{\boldsymbol{x}}(t_0)=\hat{\boldsymbol{x}}(\frac{\tau s_0}{\tau'})=\hat{\boldsymbol{y}}(s_0)$. This means that $\{\hat{\boldsymbol{x}}(t)\}_\tau$ and $\{\hat{\boldsymbol{y}}(t)\}_{\tau'}$ refer to the same OESS. Similarly, $\{\hat{\boldsymbol{x}}'(t)\}_{\tau'}$ and $\{\hat{\boldsymbol{y}}'(t)\}_{\tau}$ correspond to the same OESS (if such an OESS exists) given that $\hat{\boldsymbol{x}}'(s_0)=\hat{\boldsymbol{x}}'(\frac{\tau' t_0}{\tau})=\hat{\boldsymbol{y}}'(t_0)$.

Now, since $ \hat{\boldsymbol{y}}'(t)=(\hat{y}'_1(t), \hat{y}'_2(t), \cdots, \hat{y}'_n(t))$, we get the following:
\begin{eqnarray}
	\frac{1}{\tau}\int_{0}^{\tau}\hat{\boldsymbol{y}}'(t_0)\cdot   {\sf A}\hat{\bar{\boldsymbol{x}}}dt_0
	=\frac{1}{\tau}\int_{0}^{\tau}\sum_{i\in \text{supp}(\hat{\boldsymbol{y}}')}\hat{y}_i'(t_0)({\sf A}\hat{\bar{\boldsymbol{x}}})_idt_0.\qquad\label{eq:missed}
\end{eqnarray}
However, as shown in the \emph{Lemma} of Appendix~\ref{sec:appendix_bishop_cannings}, if $\{\hat{\boldsymbol{x}}(t)\}_\tau$ is an OESS$_\tau$ and $i\in \text{supp}(\hat{\boldsymbol{x}})$, then $({\sf A}\hat{\bar{\boldsymbol{x}}})_i=\frac{1}{\tau}\int_{0}^{\tau}\hat{\boldsymbol{x}}(t_0)\cdot{\sf A}\hat{\bar{\boldsymbol{x}}}~dt_0$. Since $\text{supp}(\hat{\boldsymbol{x}}')\subseteq\text{supp}(\hat{\boldsymbol{x}})\implies\text{supp}(\hat{\boldsymbol{y}}')\subseteq\text{supp}(\hat{\boldsymbol{x}})$, it follows that $\forall i\in \text{supp}(\hat{\boldsymbol{y}}')$, we also have $i\in \text{supp}(\hat{\boldsymbol{x}})$. Therefore, we obtain from Eq.~(\ref{eq:missed}):
\begin{eqnarray}
{\frac{1}{\tau}\int_{0}^{\tau}\hat{\boldsymbol{y}}'(t_0)\cdot   {\sf A}\hat{\bar{\boldsymbol{x}}}~dt_0}
	=\frac{1}{\tau}\int_{0}^{\tau}\hat{\boldsymbol{x}}(t_0)\cdot{\sf A}\hat{\bar{\boldsymbol{x}}}~dt_0,\label{eq:(i)}
\end{eqnarray}
But given Eq.~(\ref{eq:(i)}) and the assumption that  $\{\hat{\boldsymbol{x}}(t)\}_\tau$ is an OESS$_\tau$, the Definition 1b needs inequality~(\ref{eq:ess_stab}) to be further satisfied, i.e.,
\begin{equation}\label{eqn:OESS_step2}
	\frac{1}{\tau}\int_{0}^{\tau}\hat{\boldsymbol{x}}(t_0)\cdot {\sf A}\hat{\bar{\boldsymbol{y}}}'~dt_0>\frac{1}{\tau}\int_{0}^{\tau}\hat{\boldsymbol{y}}'(t_0)\cdot{\sf A}\hat{\bar{\boldsymbol{y}}}'~dt_0.
\end{equation}
Inequality~(\ref{eqn:OESS_step2}) can be transformed using $t_0=\tau s_0/\tau'$ into the following:
\begin{equation}\label{eqn:OESS_step3}
	\frac{1}{\tau'}\int_{0}^{\tau'}\hat{\boldsymbol{y}}(s_0)\cdot {\sf A}\hat{\bar{\boldsymbol{x}}}'~ds_0>\frac{1}{\tau'}\int_{0}^{\tau'}\hat{\boldsymbol{x}}'(s_0)\cdot{\sf A}\hat{\bar{\boldsymbol{x}}}'~ds_0,
\end{equation}
where we have made use of the obvious relation: $\hat{\bar{\boldsymbol{x}}}'=\hat{\bar{\boldsymbol{y}}}'$. Since $\{\hat{\boldsymbol{x}}'(t)\}_{\tau'}\ne \{\hat{\boldsymbol{y}}(t)\}_{\tau'}$, i.e., the time-scaling cannot convert $\{\hat{\boldsymbol{x}}(t)\}_{\tau}$ to $\{\hat{\boldsymbol{x}}'(t)\}_{\tau'}$, inequality~(\ref{eqn:OESS_step3}) directly implies $\{\hat{\boldsymbol{x}}'(t')\}_{\tau'}$ cannot be OESS, as it violates inequality~(\ref{eqn:equlibrium_1b}). Thus, the proposition stands proven. \qed
\\
\\
\noindent\textit{\textbf{Proposition~1b:} There cannot exist simultaneously an OESS $\{\hat{\boldsymbol{x}}(t)\}_{\tau}$ and
an ESS $\hat{\boldsymbol{x}}'$ such that
$\text{supp}(\hat{\boldsymbol{x}}') \subseteq
\text{supp}(\hat{\boldsymbol{x}})$.}\\
\textit{Proof:} Assume that $\{\hat{\boldsymbol{x}}(t)\}_{\tau}$ is an OESS and consider a fixed phase point $\hat{\boldsymbol{x}}'$ such that
$\text{supp}(\hat{\boldsymbol{x}}') \subseteq
\text{supp}(\hat{\boldsymbol{x}})$. Then
\begin{eqnarray}
\frac{1}{\tau}\int_{0}^{\tau}\hat{\boldsymbol{x}}(t_0)\cdot{\sf A}\hat{\bar{\boldsymbol{x}}}~dt_0&=&\sum_{i\in \text{supp}(\hat{\boldsymbol{x}}')}\hat{x}'_i\frac{1}{\tau}\int_{0}^{\tau}\hat{\boldsymbol{x}}(t_0)\cdot{\sf A}\hat{\bar{\boldsymbol{x}}}~dt_0\nonumber\\
	&=&\sum_{i\in \text{supp}(\hat{\boldsymbol{x}}')}\hat{x}_i'({\sf A}\hat{\bar{\boldsymbol{x}}})_i\nonumber\\
	&=&\frac{1}{\tau}\int_{0}^{\tau}\hat{\boldsymbol{x}}'(t_0)\cdot{\sf A}\hat{\bar{\boldsymbol{x}}}~dt_0.
\end{eqnarray}
Here, the first equality is because $\sum_{i\in \text{supp}(\hat{\boldsymbol{x}}')}\hat{x}'_i=1$, while the second one follows from the Lemma in Appendix~\ref{sec:appendix_bishop_cannings} and the assumption that $\text{supp}(\hat{\boldsymbol{x}}') \subseteq
\text{supp}(\hat{\boldsymbol{x}})$.  The last equality follows because of $\hat{\boldsymbol{x}}'(t_0)=\hat{\boldsymbol{x}}'$ being constant in time.

In view of the Definition~1b, since $\{\hat{\boldsymbol{x}}(t)\}_{\tau}$ is an OESS, it must satisfy
inequality~(\ref{eq:ess_stab}), viz., $\frac{1}{\tau}\int_{0}^{\tau}\hat{\boldsymbol{x}}(t_0)\cdot{\sf A}\hat{\bar{\boldsymbol{x}}}'~dt_0>\frac{1}{\tau}\int_{0}^{\tau}\hat{\boldsymbol{x}}'(t_0)\cdot{\sf A}\hat{\bar{\boldsymbol{x}}}'~dt_0$. Since $\hat{\boldsymbol{x}}'(t_0)$ is constant in time, this reduces to
$\hat{\bar{\boldsymbol{x}}}\cdot{\sf A}\hat{\boldsymbol{x}}'>\hat{\boldsymbol{x}}'\cdot{\sf A}\hat{\boldsymbol{x}}'$,
which violates the condition for $\hat{\boldsymbol{x}}'$ to be an ESS. Thus, the proposition is proven.
\qed
\\
\\
Two points are worth pointing out. Firstly, the simultaneous non-existence of an ESS and an OESS can also be proved
for the case $\operatorname{supp}(\hat{\boldsymbol{x}})
\subseteq \operatorname{supp}(\hat{\boldsymbol{x}}')$,
by analogous arguments. Secondly, both Propositions~1a and~1b implicitly assume that
$\hat{\bar{\boldsymbol{x}}} \neq \hat{\bar{\boldsymbol{x}}}'$; in other words, both propositions and their consequences exclude the non-generic case in which $\hat{\bar{\boldsymbol{x}}} = \hat{\bar{\boldsymbol{x}}}'$.

We find, as corollaries of the aforementioned propositions, some useful and important conclusions.
\begin{itemize}
	\item {Uniqueness of  OESS:}
	Once an $\mathrm{OESS}_{\tau}$ has been identified whose support contains all the vertices of the ($n-1$)-dimensional simplex, no other OESS can exist. In this sense, the completely interior OESS is unique. Clearly, two OESSes may coexist only on two different faces---($n-2$)-dimensional sub-simplices---of the phase space.
	\item {OESS versus ESS:} If there is any ESS, there can't be any completely interior OESS. In the presence of ESS, an OESS can exist only if it is in a sub-simplex that does not contain the ESS.
\end{itemize}

\begin{figure*}[t]
	\centering	
	\includegraphics[width=\textwidth, scale=0.50]{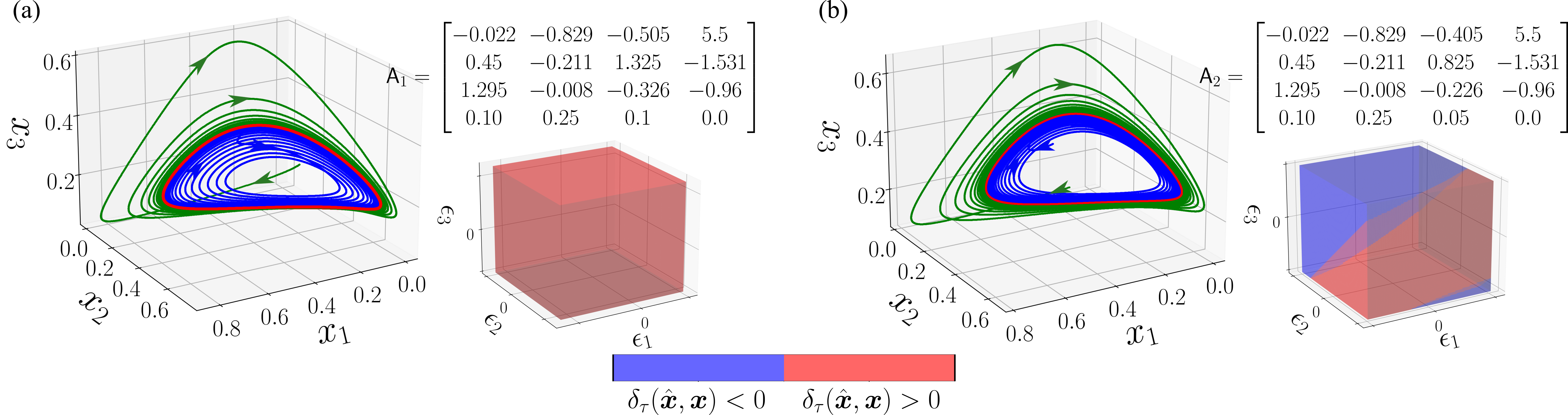}
	\caption{OESS is a stable limit cycle: A red closed curve denotes a limit cycle, and the blue and the green curves are neighbouring trajectories in the basin of attraction of the limit cycle. In subplot (a), the limit cycle in the replicator dynamics with ${\sf A}={\sf A}_1$ satisfies the OESS conditions as validated in the colour plot of $\delta_\tau(\hat{\boldsymbol{x}},\boldsymbol{x})\equiv
		\frac{1}{\tau}\int_{0}^{\tau} \hat{\boldsymbol{x}}(t_0)\cdot  {\sf A} \bar{\boldsymbol{x}}(t_0)~dt_0-\frac{1}{\tau}\int_{0}^{\tau} {\boldsymbol{x}}(t_0)\cdot  {\sf A} \bar{\boldsymbol{x}}(t_0)~dt_0$. Note $\delta_\tau(\hat{\boldsymbol{x}},\boldsymbol{x})>0$ implies that the OESS condition is satisfied. A similar subplot (b)---now with ${\sf A}={\sf A}_2$---depicts a limit cycle which is not an OESS. The triplet $(\epsilon_1,\epsilon_2,\epsilon_3)$ represents different choices of initial conditions for neighboring trajectories (see Appendix~\ref{sec:numerical}).}
	\label{fig:1}
\end{figure*}
Finally, we set to do the important step that establishes the relationship between OESS and the stability of a limit cycle within the framework of replicator dynamics. 
\section{Dynamical Stability of OESS}
\noindent\textit{\textbf{Theorem:} OESS$_{\tau}$ is a locally asymptotically stable limit cycle of period $\tau$ in replicator dynamics.}\\
\textit{Proof:}
The crucial step towards establishing the theorem is to reduce the problem of analyzing the stability of a closed orbit to that of studying the stability of a corresponding fixed point of the stroboscopic map which samples the state at intervals of length 
$\tau$. Integrating Eq.~(\ref{eq:replicator})  from $t_0-\tau$ to $t_0$, we get the stroboscopic map (see Appendix~\ref{app:sm}):
\begin{equation} \label{eq:strobos_map}
	x_i(t_0) = x_i(t_0-\tau) \cdot \exp\left(  ({\sf A}\bar{\boldsymbol{x}}(t_0))_i - \overline{\boldsymbol{x} \cdot {\sf A}\boldsymbol{x}}(t_0) \right),
\end{equation}
where $\overline{\boldsymbol{x} \cdot {\sf A}\boldsymbol{x}}(t_0)=\int_{t_0-\tau}^{t_0}\boldsymbol{x}(t) \cdot {\sf A}\boldsymbol{x}(t)~dt$. This map gives the state variables after every fixed interval $\tau$. 

Now, let us consider, $\boldsymbol{x}(t)=\boldsymbol{s}(t)+\hat{\boldsymbol{x}}(t)$, where the term $\boldsymbol{s}(t)$ captures small perturbations about the orbit $\hat{\boldsymbol{x}}(t)$, clearly $\sum_{i}s_i(t)=0$. We denote by $\mathbb{D}_x$ the domain of all nearby trajectories $\boldsymbol{x}(t)$ and by $\mathbb{D}_s$ the corresponding domain of perturbations $\boldsymbol{s}(t)$.
Substituting $x_i(t)=s_i(t)+\hat{x}_i(t)$ in Eq.~(\ref{eq:strobos_map}) we get,
\begin{eqnarray} 
	s_i(t_0)+\hat{x}_i(t_0)= \left[s_i(t_0-\tau)+\hat{x}_i(t_0-\tau)\right]\times\nonumber\qquad\qquad\\
	\exp\Big[\tau \big\{({\sf A}\hat{\bar{\boldsymbol{x}}})_i - \overline{\hat{\boldsymbol{x}}\cdot {\sf A}\hat{\boldsymbol{x}}}+({\sf A}\bar{\boldsymbol{s}}(t_0)_i-\overline{\hat{\boldsymbol x} \cdot {\sf A}{\boldsymbol s}}(t_0)\nonumber\\
~~~~-\overline{\boldsymbol{s} \cdot {\sf A}\hat{\boldsymbol{x}}}(t_0) \big\}\Big].\quad
\end{eqnarray}
On the periodic orbit, the periodicity 
condition $\hat{\boldsymbol{x}}(t_0-\tau)=\hat{\boldsymbol{x}}(t_0)$ implies [see Eq.~(\ref{eq:strobos_map})] that $\forall i$,
$({\sf A}\hat{\bar{\boldsymbol{x}}})_i
=\overline{\hat{\boldsymbol{x}}\cdot{\sf A}\hat{\boldsymbol{x}}}= c$, where $c$ is some constant. Therefore, taking terms only up to linear order in $s_i$'s, we will get the following map:
\begin{eqnarray}\label{eqn:for_map_s}
	 s_i(t_0)=s_i(t_0-\tau)+\tau \hat{x}_i(t_0-\tau)\times\nonumber\qquad\qquad\quad\\
	\Big[({\sf A}\bar{\boldsymbol{s}}(t_0))_i-\overline{\hat{\boldsymbol{x}}\cdot {\sf A}\boldsymbol{s}}(t_0)-\overline{\boldsymbol{s}\cdot {\sf A}\hat{\boldsymbol{x}}}(t_0)\Big]. 
\end{eqnarray}
Using the above expression, we are going to see that the function
	\begin{eqnarray}{\label{eq:lyapunov1}}
		V(\boldsymbol{s}(n\tau)) \equiv\frac{1}{\tau} \int_{0}^{\tau} \sum_{i\in{\rm supp}(\hat{\boldsymbol{x}})}\frac{s^2_i(t'+n\tau)}{2\hat{x}_i(t'-\tau)} ~dt'
	\end{eqnarray}
	is a Lyapunov function corresponding to the fixed point at the origin of the stroboscopic map given in Eq.~(\ref{eqn:for_map_s}). That $V(\boldsymbol{s}(n\tau)) $ is indeed an explicit function of $\boldsymbol{s}(n\tau) $ is discussed in Appendix~\ref{app:V(s)}.

One can easily check that the function $V(\boldsymbol{s}(n\tau))$ has a minimum at $\boldsymbol{s}=0$ and at the minimum, its value is zero. It has the value greater than zero at other points. Now, since the stroboscopic map~(\ref{eqn:for_map_s}) is continuous and differentiable at every point in the domain $\mathbb{D}_s$, therefore, it is Lipschitz continuous~\cite{Ruggero_2018}. Finally, to establish that $V$ is a local Lyapunov function, we need to show
\begin{equation}
	\Delta_{\tau}  V(\boldsymbol{s}) \equiv V(\boldsymbol{s}(\tau))-V(\boldsymbol{s}(0))<0~~\forall \boldsymbol{s}(0)\in\mathbb{D}_s.
\end{equation}
Considering terms only up to the first order in $\Delta_\tau s_i(t')\equiv s_i(t')-s_i(t'-\tau)$, we obtain the following,
\begin{equation}
	\Delta_{\tau}  V(\boldsymbol{s})= \frac{1}{\tau} \int_{0}^{\tau} \sum_{i\in{\rm supp}(\hat{\boldsymbol{x}})} \frac{2s_i(t')\Delta_{\tau} s_i(t')}{2\hat{x}_i(t'-\tau)}~dt',\label{eq:vmissed}
\end{equation}
where using Eq.~(\ref{eqn:for_map_s}) 
\begin{eqnarray}\label{eqn:for_delta_s}
	\Delta_\tau s_i(t')=\tau \hat{x}_i(t'-\tau)\Big[({\sf A}\bar{\boldsymbol{s}}(t'))_i-\overline{\hat{\boldsymbol{x}}\cdot {\sf A}\boldsymbol{s}}(t')-\overline{\boldsymbol{s}\cdot {\sf A}\hat{\boldsymbol{x}}}(t')\Big]. \nonumber\\
\end{eqnarray}
Substituting Eq.~(\ref{eqn:for_delta_s}) in Eq.~(\ref{eq:vmissed}), we get the following
\begin{eqnarray}
	\Delta_{\tau}  V(\boldsymbol{s})&&=\tau\frac{1}{\tau} \int_{0}^{\tau}\sum_{i\in{\rm supp}(\hat{\boldsymbol{x}})}s_i(t')~ \Big[({\sf A}\bar{\boldsymbol{s}}(t'))_i.\nonumber\\
	&&~~~~~~~~~~~~~-\overline{\hat{\boldsymbol{x}}\cdot {\sf A}\boldsymbol{s}}(t')
	-\overline{\boldsymbol{s} \cdot {\sf A}\hat{\boldsymbol{x}}}(t')\Big]~dt' \nonumber\\
	\implies\,\Delta_{\tau}  V(\boldsymbol{s})&&=\int_{0}^{\tau}  {\boldsymbol{s}}(t')\cdot {\sf A}\bar{\boldsymbol{s}}(t')~dt'.\label{eqn:delta_V(s)last}
\end{eqnarray} 
Hence if $\Delta_{\tau}  V(\boldsymbol{s})<0$ for all $\boldsymbol{s}(0)\in\mathbb{D}_s$, then $V(\boldsymbol{s})$ is formally the Lyapunov function and $\boldsymbol{s}(0)=0$ is a locally asymptotically stable fixed point; equivalently, $\hat{\boldsymbol{{x}}}(0)$ is a locally asymptotically stable fixed point of Eq.~(\ref{eq:strobos_map})~\cite{Guckenheimer1983}. Furthermore, we observe that $\hat{\boldsymbol{x}}(0)$ could be chosen as any arbitrary point on the periodic orbit $\hat{\boldsymbol{x}}(t)$ depending on the construction of the stroboscopic map. This fact implies that the limit cycle $\{\hat{\boldsymbol{x}}(t)\}_{\tau}$ is locally asymptotically stable.

Finally, let us recast Eq.~(\ref{eqn:delta_V(s)last}) in terms of $\boldsymbol{x}(t')$ and $\hat{\boldsymbol{x}}(t')$:
\begin{eqnarray}
\Delta_{\tau}  V&&=\int_{0}^{\tau}\boldsymbol{x}(t') \cdot{\sf A}\bar{\boldsymbol{x}}(t')~dt'-\int_{0}^{\tau} \hat{\boldsymbol{x}}(t') \cdot{\sf A}\bar{\boldsymbol{x}}(t')~dt'\nonumber\\
&&\phantom{=}-\int_{0}^{\tau}\boldsymbol{x}(t') \cdot{\sf A}\hat{\bar{\boldsymbol{x}}}~dt'
+\int_{0}^{\tau}\hat{\boldsymbol{x}}(t')\cdot{\sf A}\hat{\bar{\boldsymbol x}}~dt'.\label{eq:interm}
\end{eqnarray}
Since on the periodic orbit itself $({\sf A}\hat{\bar{\boldsymbol{x}}})_i=c$ $\forall i$, $\int_{0}^{\tau} \hat{\boldsymbol{x}}(t')\cdot{\sf A}\hat{\bar{\boldsymbol{x}}}~dt'=\int_{0}^{\tau} \boldsymbol{x}(t')\cdot{\sf A}\hat{\bar{\boldsymbol{x}}}~dt'=c\tau$. This when put in Eq.~(\ref{eq:interm}), the equation simplifies to
\begin{equation}
\Delta_{\tau}  V=-\int_{0}^{\tau}\hat{\boldsymbol{x}}(t')\cdot  {\sf A} \bar{\boldsymbol{x}}(t')~dt'
	+ \int_{0}^{\tau}
	\boldsymbol{x}(t')\cdot {\sf A} \bar{\boldsymbol{x}}(t')~dt'.\label{eq:interm1}
\end{equation}
As per Definition 1c~[inequality~(\ref{eq:OESS_neighbourhood})],  $\Delta_{\tau} V < 0$ in inequality~(\ref{eq:interm1}) means that $\{\hat{\boldsymbol{x}}(t)\}_{\tau}$ is an OESS, hence proving the theorem. \qed

It should be noted that the converse of the theorem may not hold true: a locally asymptotically stable limit cycle need not be an OESS. Fig.~\ref{fig:1} presents two illustrative examples (found by trial and error)---one where a stable limit cycle is OESS and another where the stable limit cycle is not an OESS.
\section{Conclusion}
In summary, in evolutionary systems where the underlying population dynamics exhibit persistent oscillations, instantaneous fitness comparisons may no longer be meaningful; instead, evolutionary stability must be assessed through some time-averaged quantities defined along the evolutionary trajectories. This perspective naturally generalizes the invasion-based framework of classical ESS theory to the theory of OESS which provides a consistent foundation for defining evolutionary stability in periodically fluctuating populations.

It must be pointed out that our paper does not define the evolutionary stability of a limit cycle via its dynamical stability---rather, the latter is a feature of the OESS definition exactly as what happens in the case of well-accepted concept of ESS. Thus, our work is inherently different from some past works related to limit cycles in eco-evolutionary dynamics~\cite{Reed1984,Lawlor1976,Grunert2021} which define evolutionary stability of limit cycle through their dynamical stability. Moreover, in these works, the existence and stability of limit cycle are ensured by invoking ecological dynamics into the evolutionary dynamics. This is not what we are interested in this paper: We have focussed solely on  evolutionary dynamics where population is at a demographic equilibrium---a situation well-described by the paradigmatic replicator dynamics whose mention is absent in the afore-cited works.

We conclude by raising one intriguing question deserving future investigation: In continuous-time replicator dynamics, how can game-theoretical evolutionary stability be associated with \emph{chaotically} oscillating populations?

\acknowledgements
V.K.D. thanks I.I.T. Kanpur (India) for the financial support through Fellowship for Academic and Research Excellence (FARE). 
\appendix

\section{Equivalence between alternative OESS definitions}\label{sec:equivalence_state}
In this section, we are going to show that Definitions 1a, 1b and 1c are equivalent.
\\
\\
\noindent\textit{\textbf{Proposition:} Definition~1a, Definition~1b and Definition~1c are equivalent.}\\
\textit{Proof:} We shall prove it in two parts.\\
Part (i): Definition~1a $\Leftrightarrow$ Definition~1b. 

\noindent First, we rewrite inequality~(\ref{eq:ess_general}) in the following form:
\begin{eqnarray}\label{eq:expansion}
	(1 - \epsilon)\left[ \frac{1}{\tau}\int_{0}^{\tau}\hat{\boldsymbol{x}}(t_0)\cdot  {\sf A} \hat{\bar{\boldsymbol{x}}}\,dt_0 - \frac{1}{\tau}\int_{0}^{\tau}\boldsymbol{x}(t_0)\cdot  {\sf A} \hat{\bar{\boldsymbol{x}}} \, dt_0 \right]
	+\nonumber\\
	\epsilon \left[\frac{1}{\tau} \int_{0}^{\tau}\hat{\boldsymbol{x}}(t_0)\cdot  {\sf A} \boldsymbol{\bar{x}}(t_0) \, dt_0-\frac{1}{\tau}\int_{0}^{\tau} \boldsymbol{x}(t_0)\cdot  {\sf A} \boldsymbol{\bar{x}}(t_0)\, dt_0 \right]
	> 0.\nonumber\\
\end{eqnarray}
One can see in the limit of a vanishing mutant fraction, $\epsilon \to 0$, this condition reduces to inequality~(\ref{eqn:equlibrium_1b}). Furthermore, if the coefficient of $(1-\epsilon)$ vanishes for some $\{\boldsymbol{x}(t)\}_{t \in [-\tau,\tau]} \ne \{\hat{\boldsymbol{x}}(t)\}_\tau$, then the coefficient of $\epsilon$ must be positive; this is precisely inequality~(\ref{eq:ess_stab}). Therefore, Definition~1a implies Definition~1b. Conversely, Definition~1b clearly implies Definition~1a, since plugging inequality~(\ref{eqn:equlibrium_1b}) and~(\ref{eq:ess_stab}) as the coefficients of $1-\epsilon$ and $\epsilon$, respectively, directly yields inequality~(\ref{eq:expansion}). We thus conclude that Definition~1a and Definition~1b are equivalent.

\noindent Part (ii): Definition~1a $\Leftrightarrow$ Definition~1c.\\
We start by assuming that the state sequence $\{\hat{\boldsymbol{x}}(t)\}_\tau$ is an $ \mathrm{OESS}$ according to Definition 1a.  Any state  $\boldsymbol{x}'(t)$ at any time $t$ that is close to $\hat{\boldsymbol{x}}(t)$ can be written as
\begin{equation}\label{eqn:neighborhood_vector}
	\boldsymbol{x}'(t)\equiv(1-\epsilon)\hat{\boldsymbol{x}}(t)+\epsilon \boldsymbol{{x}}(t),
\end{equation}
for sufficiently small $\epsilon$. It suffices to consider $\boldsymbol{x}(t)$ from the compact set $\mathbb{C}(t)\;\equiv\;\left\{\boldsymbol{x}(t)\in S_n : x_i(t)=0 \;\text{for some}\; i\in\text{supp}(\hat{\boldsymbol{x}}(t))\right\}$ [where $S_n \equiv \left\{ (x_1, \cdots, x_n) \in \mathbb{R}^n \mid x_i \geq 0, \sum_{i=1}^{n} x_i = 1 \right\}$],
which is the union of all faces of the simplex that do not contain 
$\hat{\boldsymbol{x}}(t)$ at time $t$. For all $\boldsymbol{{x}}(t)\in \mathbb{C}(t)$, Definition~1a is satisfied for $\forall\epsilon$ such that $0<\epsilon<\bar{\epsilon}(\boldsymbol{x}(t))$. We can always construct a positive continuous function $\bar{\epsilon}(\boldsymbol{x}(t))\in (0,1]$ and since $\mathbb{C}$ is compact, $\epsilon_{\min}(t)=\min\{\bar{\epsilon}(\boldsymbol{x}(t)): \boldsymbol{x}(t)\in\mathbb{C}\}$ strictly positive at any time $t$. Since $\epsilon_{\min}(t)$ is strictly positive at all $t$, the minimum $\min_t\epsilon_{\min}(t)$ is also strictly positive. Therefore Definition 1a holds for all $\epsilon<\min_t\epsilon_{\min}(t)$. Subsequently, we multiply by $\epsilon$ and add
$\frac{(1-\epsilon)}{\tau}\int_{0}^{\tau}\hat{\boldsymbol{x}}(t_0)\cdot {\sf A}  \left[ \epsilon\bar{ \boldsymbol{x}}(t_0) + (1 - \epsilon) \hat{\bar{\boldsymbol{x}}} \right]~dt_0$
to both sides of Definition 1a [inequality~(\ref{eq:ess_general})], and use Eq.~(\ref{eqn:neighborhood_vector}) to ultimately reach the following inequality:
\begin{equation}
	\frac{1}{\tau} \int_{0}^{\tau} \hat{\boldsymbol{x}}(t_0)\cdot  {\sf A} \bar{\boldsymbol{x}}'(t_0)~dt_0
	>
	\frac{1}{\tau} \int_{0}^{\tau} \boldsymbol{x}'(t_0)\cdot {\sf A} \bar{\boldsymbol{x}}'(t_0)~dt_0.
\end{equation} 
Upon relabeling $\boldsymbol{x}'(t)$ as $\boldsymbol{x}(t)$ we get inequality~(\ref{eq:OESS_neighbourhood}).
One can similarly do the steps in reverse order to prove that Definition~1c implies Definition~1a. Therefore, Definition 1a and Definition 1c are equivalent. 

Combining the conclusions from both the parts of the proofs, we conclude that all the three definitions are equivalent. \qed

\section{Proving $({\sf A}\hat{\bar{\boldsymbol{x}}})_i=\hat{\bar{\boldsymbol{x}}}\cdot{\sf A}\hat{\bar{\boldsymbol{x}}}$}\label{sec:appendix_bishop_cannings}
\noindent\textit{\textbf{Lemma:}} \textit{If $\{\hat{\boldsymbol{x}}(t)\}_\tau$ is an OESS and ${\boldsymbol{e}_i\in \text{supp}(\hat{\boldsymbol{x}})}$, then $({\sf A}\hat{\bar{\boldsymbol{x}}})_i=\frac{1}{\tau}\int_{0}^{\tau}\hat{\boldsymbol{x}}(t_0)\cdot{\sf A}\hat{\bar{\boldsymbol{x}}}~dt_0=\hat{\bar{\boldsymbol{x}}}\cdot{\sf A}\hat{\bar{\boldsymbol{x}}}$.}\\
\textit{Proof:} Since $\{\hat{\boldsymbol{x}}(t)\}_\tau$ is an OESS, therefore, from Definition 1b (inequality~(\ref{eqn:equlibrium_1b})),  \begin{equation}
	\frac{1}{\tau}\int_{0}^{\tau}\hat{\boldsymbol{x}}(t_0)\cdot{\sf A}\hat{\bar{\boldsymbol{x}}}\, dt_0\geq({\sf A}\hat{\bar{\boldsymbol{x}}})_i,\label{eq:apm}
\end{equation}
for mutant sequences such that $\{x_i(t)=1,~x_j(t)=0, \forall j\neq i\}_{t\in [-\tau,\tau]}$.

Next, we write the time average of the sequence $\{\hat{\boldsymbol{x}}(t)\}_\tau$ as follows:
\begin{equation}
	\hat{\bar{\boldsymbol{x}}}=\hat{\bar{x}}_i\boldsymbol{e}_i+(1-\hat{\bar{x}}_i)\hat{\bar{\boldsymbol{x}}}_{-i},
\end{equation}
where $\hat{\bar{\boldsymbol{x}}}_{-i}$ is the time average of a mixed state which does not include the $i$th pure state $\boldsymbol{e}_i$; in other words, $\hat{\boldsymbol{x}}_{-i}(t)$ is a mixed state in a subspace that excludes the unit vector $\boldsymbol{e}_i$.

We are going to prove the lemma by contradiction---for that let us assume
\begin{equation}\label{eqn:assumption_BC}
	\frac{1}{\tau}\int_{0}^{\tau}\hat{\boldsymbol{x}}(t_0)\cdot{\sf A}\hat{\bar{\boldsymbol{x}}}~dt_0>({\sf A}\hat{\bar{\boldsymbol{x}}})_i.
\end{equation}
Since the state sequence $\{\hat{\boldsymbol{x}}(t)\}_\tau$ is periodic, we can write
\begin{eqnarray}
	\frac{1}{\tau}\int_{0}^{\tau}\hat{\boldsymbol{x}}(t_0)\cdot{\sf A}\hat{\bar{\boldsymbol{x}}}~dt_0&&=\hat{\bar{\boldsymbol{x}}}\cdot{\sf A}\hat{\bar{\boldsymbol{x}}},\nonumber\\
	&& =
	\hat{\bar{x}}_i({\sf A}\hat{\bar{\boldsymbol{x}}})_i+(1-\hat{\bar{x}}_i)\hat{\bar{\boldsymbol{x}}}_{-i}\cdot{\sf A}\hat{\bar{\boldsymbol{x}}}\nonumber\\
	&&<\hat{\bar{x}}_i \hat{\bar{\boldsymbol{x}}}\cdot{\sf A}\hat{\bar{\boldsymbol{x}}}+(1-\hat{\bar{x}}_i)\hat{\bar{\boldsymbol{x}}}_{-i}\cdot{\sf A}\hat{\bar{\boldsymbol{x}}}.\qquad
\end{eqnarray}
In the last inequality, we have used assumption (\ref{eqn:assumption_BC}). Rearranging the terms automatically implies
\begin{equation}
	\frac{1}{\tau}\int_{0}^{\tau}\hat{\boldsymbol{x}}(t_0)\cdot{\sf A}\hat{\bar{\boldsymbol{x}}}~dt_0<\frac{1}{\tau}\int_{0}^{\tau}\hat{\boldsymbol{x}}_{-i}(t_0)\cdot{\sf A}\hat{\bar{\boldsymbol{x}}}~dt_0,
\end{equation}
but this is not possible since $\{\hat{\boldsymbol{x}}(t)\}_\tau$ is an OESS; hence our assumption, inequality~(\ref{eqn:assumption_BC}), is incorrect. In the light of Eq.~(\ref{eq:apm}), this straightaway implies that $({\sf A}\hat{\bar{\boldsymbol{x}}})_i=\frac{1}{\tau}\int_{0}^{\tau}\hat{\boldsymbol{x}}(t_0)\cdot{\sf A}\hat{\bar{\boldsymbol{x}}}~dt_0$.\qed

We observe that this lemma reminds one of the Bishop--Cannings theorem~\cite{Bishop1978,smith_1982} given in the context of ESS. 

\section{Stoboscopic Map}\label{app:sm}
Note that although we write $t_0$ in the arguments of 
$\overline{\boldsymbol{x} \cdot {\sf A}\boldsymbol{x}}(t_0)$ and 
$({\sf A}\bar{\boldsymbol{x}}(t_0))_i$ as a shorthand in the RHS of 
\begin{equation} 
	x_i(t_0) = x_i(t_0-\tau) \cdot \exp\left(  ({\sf A}\bar{\boldsymbol{x}}(t_0))_i - \overline{\boldsymbol{x} \cdot {\sf A}\boldsymbol{x}}(t_0) \right),\label{eq:strobos_map1}
\end{equation}
(i.e., Eq.~(\ref{eq:strobos_map}) of the main text), they actually depend explicitly on $\boldsymbol{x}(t_0 - \tau)$ and $\tau$, as we explain 
below to ensure that Eq.~(\ref{eq:strobos_map}) indeed defines a stroboscopic map.

Let the solution of Eq.~(\ref{eq:replicator}) at time $t$ with initial condition 
$\boldsymbol{x}(0)$ be denoted by
\begin{equation}
	\boldsymbol{x}(t-0) = \boldsymbol{f}( t-0;\boldsymbol{x}(0)).
\end{equation}
Since $\boldsymbol{f}( t-0;\boldsymbol{x}(0))$ is the flow generated by an autonomous dynamical
system, it satisfies the property~\cite{wiggins_2003},
\begin{equation}
	f_i\left((t-t')+ (t'-0);\boldsymbol{x}(0)\right)
	= f_i\!\left( (t-t'); \boldsymbol{x}(t'-0)\right),
\end{equation}
for every $i \in \{1,\cdots,n\}$.  
Using this property, we obtain
\begin{eqnarray}
	x_i((n+1)\tau-0)
	&&= f_i\left(\{(n+1)\tau-n\tau\}+\{n\tau-0\};\boldsymbol{x}(0)\right)\nonumber\\
	&&= f_i\!\left(\tau; \boldsymbol{x}(n\tau)\right).
\end{eqnarray}
This immediately gives,
\begin{equation}
	x_i((n+1)\tau)
	= f_i\!\left(\tau;\,\boldsymbol{x}(n\tau)\right).
\end{equation}
Substituting $n\tau = t_0 - \tau$, we finally obtain
\begin{eqnarray}\label{eq:2-new}
	x_i(t_0)
	= f_i\!\left(\tau;\,\boldsymbol{x}(t_0-\tau)\right)
	= x_i(t_0-\tau)\,
	\frac{
		f_i\!\left(\tau;\,\boldsymbol{x}(t_0-\tau)\right)
	}{
		x_i(t_0-\tau)
	}.~~~\nonumber\\
\end{eqnarray}
Comparing Eq.~(\ref{eq:2-new}) with Eq.~(\ref{eq:strobos_map1}) we get,
\begin{equation}
	f_i(\tau;\boldsymbol{x}(t_0-\tau))
	= x_i(t_0-\tau) \cdot \exp\Big(  ({\sf A}\bar{\boldsymbol{x}}(t_0))_i - \overline{\boldsymbol{x} \cdot {\sf A}\boldsymbol{x}}(t_0) \Big),
\end{equation}
which shows that the RHS of Eq.~(\ref{eq:strobos_map1}) depends only on the state $\boldsymbol{x}(t_0-\tau)$ and the parameter $\tau$.

\section{Explicit dependence of $V$ on state}
\label{app:V(s)}
We begin by writing the dynamical equation for the perturbation 
$\boldsymbol{s}(t)=\boldsymbol{x}(t)-\hat{\boldsymbol{x}}(t)$, which is obtained directly from the replicator equation. This yields:
\begin{subequations}\label{eq:pertubation}
	\begin{eqnarray}
		\frac{d\hat{x}_i(t)}{dt}&&=\hat{x}_i(t)\Big[({\sf A}{\hat{\boldsymbol{x}}}(t))_i-{\hat{\boldsymbol{x}}(t)\cdot {\sf A}{\hat{\boldsymbol{x}}}}(t)\Big].\label{eq:pertubation1}\\
		\frac{ds_i(t)}{dt}
		&&=(\hat{x}_i(t)+s_i(t)) 
		\Big[({\sf A}{\boldsymbol{s}}(t))_i\nonumber\\
		&&-{\hat{\boldsymbol{x}}(t)\cdot {\sf A}\boldsymbol{s}}(t)-{\boldsymbol{s}(t)\cdot {\sf A}\hat{\boldsymbol{x}}}(t)-{\boldsymbol{s}(t)\cdot {\sf A}{\boldsymbol{s}}}(t)\Big]\nonumber\\
		&&+s_i\Big[({\sf A}{\hat{\boldsymbol{x}}}(t))_i-{\hat{\boldsymbol{x}}(t)\cdot {\sf A}{\hat{\boldsymbol{x}}}}(t)\Big].
	\end{eqnarray}
\end{subequations}
As the limit cycle solution of ~(\ref{eq:pertubation1}),---$\hat{\boldsymbol{x}}(t)$ is assumed to be formally known, the solution for $s_i(t)$  takes the following form: $s_i(t)=g_i(t; \boldsymbol{s}(0),\hat{\boldsymbol{x}}(0))$. Since $g_i$ is a component of the vector flow of an autonomous dynamical system, it satisfies the following property~\cite{wiggins_2003}:
\begin{eqnarray}\label{eq:vs_step1}
	&&g_i((t-t_1)+(t_1-0); \boldsymbol{s}(0) ,\hat{\boldsymbol{x}}(0))\nonumber\\=&&g_i((t-t_1); \boldsymbol{s}(t_1-0),\hat{\boldsymbol{x}}(t_1-0)).\qquad
\end{eqnarray} 
Therefore, $s_i(n\tau+t')=g_i((t'); \boldsymbol{s}(n\tau),\hat{\boldsymbol{x}}(n\tau))=g_i((t'); \boldsymbol{s}(n\tau),\hat{\boldsymbol{x}}(0))$. Hence, we can write the expression of $V(\boldsymbol{s}(t))$ from Eq.~(\ref{eq:lyapunov1}) of the main text in the following equivalent form:
\begin{eqnarray}
	V(\boldsymbol{s}(n\tau)) \equiv\frac{1}{\tau} \int_{0}^{\tau} \sum_{i\in{\rm supp}(\hat{\boldsymbol{x}})}\frac{g^2_i(t'; \boldsymbol{s}(n\tau),\hat{\boldsymbol{x}}(0))}{2\hat{x}_i(t'-\tau)} ~dt'.~~~~
\end{eqnarray}
This expression showcases the explicit dependence of $V(\boldsymbol{s}(n\tau))$ on 
$\boldsymbol{s}(n\tau)$.

\section{Examples}\label{sec:numerical}
We have demonstrated that OESS$_\tau$ implies the existence of a stable limit cycle in the replicator dynamics. In this section, we provide numerical examples for the scenarios (i) where a limit cycle satisfies the OESS condition and (ii) the scenarios in which a stable limit cycle violate the OESS condition. For numerical verification, we define the following short hand notation:
\begin{eqnarray}\label{OESS_numerical}
	\delta_\tau(\hat{\boldsymbol{x}},\boldsymbol{x}) :=
	\frac{1}{\tau}\int_{0}^{\tau} \hat{\boldsymbol{x}}(t_0)\cdot  {\sf A} \bar{\boldsymbol{x}}(t_0)~dt_0\nonumber\\
	-\frac{1}{\tau}\int_{0}^{\tau} {\boldsymbol{x}}(t_0)\cdot  {\sf A} \bar{\boldsymbol{x}}(t_0)~dt_0.
\end{eqnarray}
In this notation, the OESS condition simplifies to $\delta_\tau(\hat{\boldsymbol{x}},\boldsymbol{x})>0$ for all $\{\boldsymbol{{x}}(t)\}_{t \in [-\tau,\tau]}$ in the deleted neighbourhood of $\{\hat{\boldsymbol{x}}(t)\}$.

\begin{enumerate}
	\item[(i)] For the following population game matrix corresponding to subplot~(a) of Fig.~\ref{fig:1} of the main text, the OESS condition holds for all neighbouring trajectories of the numerically found limit cycle:
	\[ {\sf A}=
	\begin{bmatrix}
		- 0.022 & - 0.829 & - 0.505 & 5.5 \\
		0.45 & - 0.211 & 1.325 & - 1.531 \\
		1.295 & - 0.008 & -0.326 & - 0.96 \\
		0.10 & 0.25 & 0.1 & 0.0
	\end{bmatrix}
	\] 
	This game has no pure Nash equilibrium, and the mixed Nash equilibrium is given by $\boldsymbol{x}^*\approx(0.256, 0.412, 0.220, 0.112)$. To verify whether $\boldsymbol{x}^*$ satisfies the ESS condition, consider a small perturbation in its neighbourhood: $\boldsymbol{x}=(0.256-\epsilon_1, 0.412-\epsilon_2, 0.220-\epsilon_3, 0.112+\epsilon_1+\epsilon_2+\epsilon_3)$. The ESS condition implies:
	\begin{eqnarray}
		\boldsymbol{x}^*\cdot{\sf A}\boldsymbol{x} - \boldsymbol{x}\cdot{\sf A} \boldsymbol{x}= -1.07 \epsilon_2 ^2+\epsilon_2  (-3.458 \epsilon_3 +4.698 \epsilon_1)\nonumber\\
		-0.534 \epsilon_3^2-0.002 \epsilon_3 +5.622 \epsilon_1 ^2+(3.95 \epsilon_3) \epsilon_1.\qquad
	\end{eqnarray} 
	One can verify that this inequality is violated for $\epsilon_1 = 0$ and $\epsilon_2 = \epsilon_3 = 0.01$, implying that $\boldsymbol{x}^*$ is not an ESS. Therefore, a possibility of existence of OESS is there and it is found to exist.
	
	\item[(ii)] For the following population game matrix corresponding to subplot~(b) of Fig.~\ref{fig:1} of the main text, the OESS condition is violated for some neighbouring trajectories of the numerically found limit cycle:
	\[ {\sf A}=
	\begin{bmatrix}
		- 0.022 & - 0.829 & - 0.405 & 5.5 \\
		0.45 & - 0.211 & 0.825 & - 1.531 \\
		1.295 & - 0.008 &  -0.226 & - 0.96 \\
		0.10 & 0.25 & 0.05 & 0.0
	\end{bmatrix}
	\] 
	Again, no pure Nash equilibrium exists, and the mixed Nash equilibrium is $\boldsymbol{x}^*\approx(0.230, 0.359, 0.309, 0.102)$. For a nearby point $\boldsymbol{x}=(0.230-\epsilon_1, 0.359-\epsilon_2, 0.309-\epsilon_3, 0.102+\epsilon_1+\epsilon_2+\epsilon_3)$, we obtain:
	\begin{eqnarray}
		\boldsymbol{x}^*\cdot{\sf A}\boldsymbol{x} - \boldsymbol{x}\cdot{\sf A} \boldsymbol{x} =-1.07 \epsilon_2^2+\epsilon_2 (-3.008\epsilon_3 +4.698 \epsilon_1)\nonumber\\
		-0.684 \epsilon_3 ^2 +5.622 \epsilon_1 ^2+3.8 \epsilon_3\epsilon_1.\qquad
	\end{eqnarray} 
	This condition is violated when $\epsilon_1 = 0$ and $\epsilon_2 = \epsilon_3 = 0.01$, indicating that the mixed Nash equilibrium is not an ESS. Therefore, a possibility of existence of OESS is there but in this case it does not exist.
\end{enumerate}

\bibliography{Chakraborty_etal_bibliography}

\begin{thebibliography}{28}%
\makeatletter
\providecommand \@ifxundefined [1]{%
 \@ifx{#1\undefined}
}%
\providecommand \@ifnum [1]{%
 \ifnum #1\expandafter \@firstoftwo
 \else \expandafter \@secondoftwo
 \fi
}%
\providecommand \@ifx [1]{%
 \ifx #1\expandafter \@firstoftwo
 \else \expandafter \@secondoftwo
 \fi
}%
\providecommand \natexlab [1]{#1}%
\providecommand \enquote  [1]{``#1''}%
\providecommand \bibnamefont  [1]{#1}%
\providecommand \bibfnamefont [1]{#1}%
\providecommand \citenamefont [1]{#1}%
\providecommand \href@noop [0]{\@secondoftwo}%
\providecommand \href [0]{\begingroup \@sanitize@url \@href}%
\providecommand \@href[1]{\@@startlink{#1}\@@href}%
\providecommand \@@href[1]{\endgroup#1\@@endlink}%
\providecommand \@sanitize@url [0]{\catcode `\\12\catcode `\$12\catcode
  `\&12\catcode `\#12\catcode `\^12\catcode `\_12\catcode `\%12\relax}%
\providecommand \@@startlink[1]{}%
\providecommand \@@endlink[0]{}%
\providecommand \url  [0]{\begingroup\@sanitize@url \@url }%
\providecommand \@url [1]{\endgroup\@href {#1}{\urlprefix }}%
\providecommand \urlprefix  [0]{URL }%
\providecommand \Eprint [0]{\href }%
\providecommand \doibase [0]{https://doi.org/}%
\providecommand \selectlanguage [0]{\@gobble}%
\providecommand \bibinfo  [0]{\@secondoftwo}%
\providecommand \bibfield  [0]{\@secondoftwo}%
\providecommand \translation [1]{[#1]}%
\providecommand \BibitemOpen [0]{}%
\providecommand \bibitemStop [0]{}%
\providecommand \bibitemNoStop [0]{.\EOS\space}%
\providecommand \EOS [0]{\spacefactor3000\relax}%
\providecommand \BibitemShut  [1]{\csname bibitem#1\endcsname}%
\let\auto@bib@innerbib\@empty
\bibitem [{\citenamefont {Maynard~Smith}\ and\ \citenamefont
  {Price}(1973)}]{SMITH1973nature}%
  \BibitemOpen
  \bibfield  {author} {\bibinfo {author} {\bibfnamefont {J.}~\bibnamefont
  {Maynard~Smith}}\ and\ \bibinfo {author} {\bibfnamefont {G.~R.}\ \bibnamefont
  {Price}},\ }\bibfield  {title} {\bibinfo {title} {The logic of animal
  conflict},\ }\href {https://doi.org/10.1038/246015a0} {\bibfield  {journal}
  {\bibinfo  {journal} {Nature}\ }\textbf {\bibinfo {volume} {246}},\ \bibinfo
  {pages} {15} (\bibinfo {year} {1973})}\BibitemShut {NoStop}%
\bibitem [{\citenamefont {Maynard~Smith}(1982)}]{smith_1982}%
  \BibitemOpen
  \bibfield  {author} {\bibinfo {author} {\bibfnamefont {J.}~\bibnamefont
  {Maynard~Smith}},\ }\href@noop {} {\emph {\bibinfo {title} {Evolution and the
  Theory of Games}}}\ (\bibinfo  {publisher} {Cambridge University Press},\
  \bibinfo {address} {Cambridge},\ \bibinfo {year} {1982})\BibitemShut
  {NoStop}%
\bibitem [{\citenamefont {Thomas}(1985)}]{Thomas1985jmb}%
  \BibitemOpen
  \bibfield  {author} {\bibinfo {author} {\bibfnamefont {B.}~\bibnamefont
  {Thomas}},\ }\bibfield  {title} {\bibinfo {title} {On evolutionarily stable
  sets},\ }\bibfield  {journal} {\bibinfo  {journal} {Journal of Mathematical
  Biology}\ }\textbf {\bibinfo {volume} {22}},\ \href
  {https://doi.org/10.1007/bf00276549} {10.1007/bf00276549} (\bibinfo {year}
  {1985})\BibitemShut {NoStop}%
\bibitem [{\citenamefont {Dawkins}(1989)}]{dawkins1989selfish}%
  \BibitemOpen
  \bibfield  {author} {\bibinfo {author} {\bibfnamefont {R.}~\bibnamefont
  {Dawkins}},\ }\href@noop {} {\emph {\bibinfo {title} {The Selfish Gene}}}\
  (\bibinfo  {publisher} {Oxford University Press},\ \bibinfo {address}
  {Oxford},\ \bibinfo {year} {1989})\BibitemShut {NoStop}%
\bibitem [{\citenamefont {Taylor}\ and\ \citenamefont
  {Jonker}(1978)}]{Taylor1978mb}%
  \BibitemOpen
  \bibfield  {author} {\bibinfo {author} {\bibfnamefont {P.~D.}\ \bibnamefont
  {Taylor}}\ and\ \bibinfo {author} {\bibfnamefont {L.~B.}\ \bibnamefont
  {Jonker}},\ }\bibfield  {title} {\bibinfo {title} {Evolutionary stable
  strategies and game dynamics},\ }\href
  {https://doi.org/10.1016/0025-5564(78)90077-9} {\bibfield  {journal}
  {\bibinfo  {journal} {Mathematical Biosciences}\ }\textbf {\bibinfo {volume}
  {40}},\ \bibinfo {pages} {145} (\bibinfo {year} {1978})}\BibitemShut
  {NoStop}%
\bibitem [{\citenamefont {Schuster}\ and\ \citenamefont
  {Sigmund}(1983)}]{Schuster1983jtb}%
  \BibitemOpen
  \bibfield  {author} {\bibinfo {author} {\bibfnamefont {P.}~\bibnamefont
  {Schuster}}\ and\ \bibinfo {author} {\bibfnamefont {K.}~\bibnamefont
  {Sigmund}},\ }\bibfield  {title} {\bibinfo {title} {Replicator dynamics},\
  }\href {https://doi.org/10.1016/0022-5193(83)90445-9} {\bibfield  {journal}
  {\bibinfo  {journal} {Journal of Theoretical Biology}\ }\textbf {\bibinfo
  {volume} {100}},\ \bibinfo {pages} {533} (\bibinfo {year}
  {1983})}\BibitemShut {NoStop}%
\bibitem [{\citenamefont {Spencer}(1864)}]{Spencer1864}%
  \BibitemOpen
  \bibfield  {author} {\bibinfo {author} {\bibfnamefont {H.}~\bibnamefont
  {Spencer}},\ }\href@noop {} {\emph {\bibinfo {title} {The Principles of
  Biology}}},\ Vol.~\bibinfo {volume} {1}\ (\bibinfo  {publisher} {Williams and
  Norgate},\ \bibinfo {address} {London},\ \bibinfo {year} {1864})\BibitemShut
  {NoStop}%
\bibitem [{\citenamefont {Darwin}(1869)}]{Darwin1869}%
  \BibitemOpen
  \bibfield  {author} {\bibinfo {author} {\bibfnamefont {C.}~\bibnamefont
  {Darwin}},\ }\href@noop {} {\emph {\bibinfo {title} {On the Origin of Species
  by Means of Natural Selection, or the Preservation of Favoured Races in the
  Struggle for Life}}},\ \bibinfo {edition} {5th}\ ed.\ (\bibinfo  {publisher}
  {John Murray},\ \bibinfo {address} {London},\ \bibinfo {year}
  {1869})\BibitemShut {NoStop}%
\bibitem [{\citenamefont {Hori}(1993)}]{Hori1993science}%
  \BibitemOpen
  \bibfield  {author} {\bibinfo {author} {\bibfnamefont {M.}~\bibnamefont
  {Hori}},\ }\bibfield  {title} {\bibinfo {title} {Frequency-dependent natural
  selection in the handedness of scale-eating cichlid fish},\ }\href
  {https://doi.org/10.1126/science.260.5105.216} {\bibfield  {journal}
  {\bibinfo  {journal} {Science}\ }\textbf {\bibinfo {volume} {260}},\ \bibinfo
  {pages} {216} (\bibinfo {year} {1993})}\BibitemShut {NoStop}%
\bibitem [{\citenamefont {Sinervo}\ and\ \citenamefont
  {Lively}(1996)}]{Sinervo1996nature}%
  \BibitemOpen
  \bibfield  {author} {\bibinfo {author} {\bibfnamefont {B.}~\bibnamefont
  {Sinervo}}\ and\ \bibinfo {author} {\bibfnamefont {C.~M.}\ \bibnamefont
  {Lively}},\ }\bibfield  {title} {\bibinfo {title} {The
  rock–paper–scissors game and the evolution of alternative male
  strategies},\ }\href {https://doi.org/10.1038/380240a0} {\bibfield  {journal}
  {\bibinfo  {journal} {Nature}\ }\textbf {\bibinfo {volume} {380}},\ \bibinfo
  {pages} {240} (\bibinfo {year} {1996})}\BibitemShut {NoStop}%
\bibitem [{\citenamefont {Sinervo}\ \emph {et~al.}(2000)\citenamefont
  {Sinervo}, \citenamefont {Svensson},\ and\ \citenamefont
  {Comendant}}]{Sinervo2000nature}%
  \BibitemOpen
  \bibfield  {author} {\bibinfo {author} {\bibfnamefont {B.}~\bibnamefont
  {Sinervo}}, \bibinfo {author} {\bibfnamefont {E.}~\bibnamefont {Svensson}},\
  and\ \bibinfo {author} {\bibfnamefont {T.}~\bibnamefont {Comendant}},\
  }\bibfield  {title} {\bibinfo {title} {Density cycles and an offspring
  quantity and quality game driven by natural selection},\ }\href
  {https://doi.org/10.1038/35023149} {\bibfield  {journal} {\bibinfo  {journal}
  {Nature}\ }\textbf {\bibinfo {volume} {406}},\ \bibinfo {pages} {985}
  (\bibinfo {year} {2000})}\BibitemShut {NoStop}%
\bibitem [{\citenamefont {Kirkup}\ and\ \citenamefont
  {Riley}(2004)}]{Kirkup2004}%
  \BibitemOpen
  \bibfield  {author} {\bibinfo {author} {\bibfnamefont {B.~C.}\ \bibnamefont
  {Kirkup}}\ and\ \bibinfo {author} {\bibfnamefont {M.~A.}\ \bibnamefont
  {Riley}},\ }\bibfield  {title} {\bibinfo {title} {Antibiotic-mediated
  antagonism leads to a bacterial game of rock–paper–scissors in vivo},\
  }\href {https://doi.org/10.1038/nature02429} {\bibfield  {journal} {\bibinfo
  {journal} {Nature}\ }\textbf {\bibinfo {volume} {428}},\ \bibinfo {pages}
  {412–414} (\bibinfo {year} {2004})}\BibitemShut {NoStop}%
\bibitem [{\citenamefont {Lotka}(1920)}]{Lotka1920pnas}%
  \BibitemOpen
  \bibfield  {author} {\bibinfo {author} {\bibfnamefont {A.~J.}\ \bibnamefont
  {Lotka}},\ }\bibfield  {title} {\bibinfo {title} {Analytical note on certain
  rhythmic relations in organic systems},\ }\href
  {https://doi.org/10.1073/pnas.6.7.410} {\bibfield  {journal} {\bibinfo
  {journal} {Proceedings of the National Academy of Sciences}\ }\textbf
  {\bibinfo {volume} {6}},\ \bibinfo {pages} {410} (\bibinfo {year}
  {1920})}\BibitemShut {NoStop}%
\bibitem [{\citenamefont {Peterson}\ \emph {et~al.}(1984)\citenamefont
  {Peterson}, \citenamefont {Page},\ and\ \citenamefont
  {Dodge}}]{Peterson1984science}%
  \BibitemOpen
  \bibfield  {author} {\bibinfo {author} {\bibfnamefont {R.~O.}\ \bibnamefont
  {Peterson}}, \bibinfo {author} {\bibfnamefont {R.~E.}\ \bibnamefont {Page}},\
  and\ \bibinfo {author} {\bibfnamefont {K.~M.}\ \bibnamefont {Dodge}},\
  }\bibfield  {title} {\bibinfo {title} {Wolves, moose, and the allometry of
  population cycles},\ }\href {https://doi.org/10.1126/science.224.4655.1350}
  {\bibfield  {journal} {\bibinfo  {journal} {Science}\ }\textbf {\bibinfo
  {volume} {224}},\ \bibinfo {pages} {1350} (\bibinfo {year}
  {1984})}\BibitemShut {NoStop}%
\bibitem [{\citenamefont {Fauteux}\ \emph {et~al.}(2015)\citenamefont
  {Fauteux}, \citenamefont {Gauthier},\ and\ \citenamefont
  {Berteaux}}]{Fauteux2015jae}%
  \BibitemOpen
  \bibfield  {author} {\bibinfo {author} {\bibfnamefont {D.}~\bibnamefont
  {Fauteux}}, \bibinfo {author} {\bibfnamefont {G.}~\bibnamefont {Gauthier}},\
  and\ \bibinfo {author} {\bibfnamefont {D.}~\bibnamefont {Berteaux}},\
  }\bibfield  {title} {\bibinfo {title} {Seasonal demography of a cyclic
  lemming population in the {C}anadian {A}rctic},\ }\href
  {https://doi.org/10.1111/1365-2656.12385} {\bibfield  {journal} {\bibinfo
  {journal} {Journal of Animal Ecology}\ }\textbf {\bibinfo {volume} {84}},\
  \bibinfo {pages} {1412} (\bibinfo {year} {2015})}\BibitemShut {NoStop}%
\bibitem [{\citenamefont {Hofbauer}\ and\ \citenamefont
  {Sigmund}(1998)}]{hofbauer_book}%
  \BibitemOpen
  \bibfield  {author} {\bibinfo {author} {\bibfnamefont {J.}~\bibnamefont
  {Hofbauer}}\ and\ \bibinfo {author} {\bibfnamefont {K.}~\bibnamefont
  {Sigmund}},\ }\href
  {http://www.amazon.ca/exec/obidos/redirect?tag=citeulike04-20{\&}path=ASIN/052162570X}
  {\emph {\bibinfo {title} {Evolutionary Games and Population Dynamics}}}\
  (\bibinfo  {publisher} {{Cambridge University Press}},\ \bibinfo {year}
  {1998})\BibitemShut {NoStop}%
\bibitem [{\citenamefont {Mukhopadhyay}\ and\ \citenamefont
  {Chakraborty}(2020{\natexlab{a}})}]{Mukhopadhyay2020jtb}%
  \BibitemOpen
  \bibfield  {author} {\bibinfo {author} {\bibfnamefont {A.}~\bibnamefont
  {Mukhopadhyay}}\ and\ \bibinfo {author} {\bibfnamefont {S.}~\bibnamefont
  {Chakraborty}},\ }\bibfield  {title} {\bibinfo {title} {Periodic orbit can be
  evolutionarily stable: Case study of discrete replicator dynamics},\ }\href
  {https://doi.org/10.1016/j.jtbi.2020.110288} {\bibfield  {journal} {\bibinfo
  {journal} {Journal of Theoretical Biology}\ }\textbf {\bibinfo {volume}
  {497}},\ \bibinfo {pages} {110288} (\bibinfo {year}
  {2020}{\natexlab{a}})}\BibitemShut {NoStop}%
\bibitem [{\citenamefont {Mukhopadhyay}\ and\ \citenamefont
  {Chakraborty}(2020{\natexlab{b}})}]{Mukhopadhyay2020chaos}%
  \BibitemOpen
  \bibfield  {author} {\bibinfo {author} {\bibfnamefont {A.}~\bibnamefont
  {Mukhopadhyay}}\ and\ \bibinfo {author} {\bibfnamefont {S.}~\bibnamefont
  {Chakraborty}},\ }\bibfield  {title} {\bibinfo {title} {Deciphering chaos in
  evolutionary games},\ }\bibfield  {journal} {\bibinfo  {journal} {Chaos: An
  Interdisciplinary Journal of Nonlinear Science}\ }\textbf {\bibinfo {volume}
  {30}},\ \href {https://doi.org/10.1063/5.0029480} {10.1063/5.0029480}
  (\bibinfo {year} {2020}{\natexlab{b}})\BibitemShut {NoStop}%
\bibitem [{\citenamefont {Bhattacharjee}\ \emph {et~al.}(2023)\citenamefont
  {Bhattacharjee}, \citenamefont {Dubey}, \citenamefont {Mukhopadhyay},\ and\
  \citenamefont {Chakraborty}}]{Bhattacharjee2023pre}%
  \BibitemOpen
  \bibfield  {author} {\bibinfo {author} {\bibfnamefont {S.}~\bibnamefont
  {Bhattacharjee}}, \bibinfo {author} {\bibfnamefont {V.~K.}\ \bibnamefont
  {Dubey}}, \bibinfo {author} {\bibfnamefont {A.}~\bibnamefont
  {Mukhopadhyay}},\ and\ \bibinfo {author} {\bibfnamefont {S.}~\bibnamefont
  {Chakraborty}},\ }\bibfield  {title} {\bibinfo {title} {Periodic orbits in
  deterministic discrete-time evolutionary game dynamics: An
  information-theoretic perspective},\ }\bibfield  {journal} {\bibinfo
  {journal} {Physical Review E}\ }\textbf {\bibinfo {volume} {107}},\ \href
  {https://doi.org/10.1103/physreve.107.064405} {10.1103/physreve.107.064405}
  (\bibinfo {year} {2023})\BibitemShut {NoStop}%
\bibitem [{\citenamefont {Dubey}\ \emph {et~al.}(2024)\citenamefont {Dubey},
  \citenamefont {Chakraborty},\ and\ \citenamefont
  {Chakraborty}}]{Dubey2024dg}%
  \BibitemOpen
  \bibfield  {author} {\bibinfo {author} {\bibfnamefont {V.~K.}\ \bibnamefont
  {Dubey}}, \bibinfo {author} {\bibfnamefont {S.}~\bibnamefont {Chakraborty}},\
  and\ \bibinfo {author} {\bibfnamefont {S.}~\bibnamefont {Chakraborty}},\
  }\bibfield  {title} {\bibinfo {title} {Oscillatory equilibrium in asymmetric
  evolutionary games: Generalizing evolutionarily stable strategy},\ }\bibfield
   {journal} {\bibinfo  {journal} {Dynamic Games and Applications}\ }\href
  {https://doi.org/10.1007/s13235-024-00606-2} {10.1007/s13235-024-00606-2}
  (\bibinfo {year} {2024})\BibitemShut {NoStop}%
\bibitem [{\citenamefont {Hofbauer}(1981)}]{Hofbauer1981}%
  \BibitemOpen
  \bibfield  {author} {\bibinfo {author} {\bibfnamefont {J.}~\bibnamefont
  {Hofbauer}},\ }\bibfield  {title} {\bibinfo {title} {On the occurrence of
  limit cycles in the volterra-lotka equation},\ }\href
  {https://doi.org/10.1016/0362-546x(81)90059-6} {\bibfield  {journal}
  {\bibinfo  {journal} {Nonlinear Analysis: Theory, Methods \& Applications}\
  }\textbf {\bibinfo {volume} {5}},\ \bibinfo {pages} {1003–1007} (\bibinfo
  {year} {1981})}\BibitemShut {NoStop}%
\bibitem [{\citenamefont {Bof}\ \emph {et~al.}(2018)\citenamefont {Bof},
  \citenamefont {Carli},\ and\ \citenamefont {Schenato}}]{Ruggero_2018}%
  \BibitemOpen
  \bibfield  {author} {\bibinfo {author} {\bibfnamefont {N.}~\bibnamefont
  {Bof}}, \bibinfo {author} {\bibfnamefont {R.}~\bibnamefont {Carli}},\ and\
  \bibinfo {author} {\bibfnamefont {L.}~\bibnamefont {Schenato}},\ }\href
  {https://doi.org/10.48550/ARXIV.1809.05289} {\bibinfo {title} {Lyapunov
  theory for discrete time systems}} (\bibinfo {year} {2018})\BibitemShut
  {NoStop}%
\bibitem [{\citenamefont {Guckenheimer}\ and\ \citenamefont
  {Holmes}(1983)}]{Guckenheimer1983}%
  \BibitemOpen
  \bibfield  {author} {\bibinfo {author} {\bibfnamefont {J.}~\bibnamefont
  {Guckenheimer}}\ and\ \bibinfo {author} {\bibfnamefont {P.}~\bibnamefont
  {Holmes}},\ }\href {https://doi.org/10.1007/978-1-4612-1140-2} {\emph
  {\bibinfo {title} {Nonlinear Oscillations, Dynamical Systems, and
  Bifurcations of Vector Fields}}}\ (\bibinfo  {publisher} {Springer New
  York},\ \bibinfo {year} {1983})\BibitemShut {NoStop}%
\bibitem [{\citenamefont {Reed}\ and\ \citenamefont
  {Stenseth}(1984)}]{Reed1984}%
  \BibitemOpen
  \bibfield  {author} {\bibinfo {author} {\bibfnamefont {J.}~\bibnamefont
  {Reed}}\ and\ \bibinfo {author} {\bibfnamefont {N.~C.}\ \bibnamefont
  {Stenseth}},\ }\bibfield  {title} {\bibinfo {title} {On evolutionarily stable
  strategies},\ }\href {https://doi.org/10.1016/s0022-5193(84)80075-2}
  {\bibfield  {journal} {\bibinfo  {journal} {Journal of Theoretical Biology}\
  }\textbf {\bibinfo {volume} {108}},\ \bibinfo {pages} {491–508} (\bibinfo
  {year} {1984})}\BibitemShut {NoStop}%
\bibitem [{\citenamefont {Lawlor}\ and\ \citenamefont
  {Smith}(1976)}]{Lawlor1976}%
  \BibitemOpen
  \bibfield  {author} {\bibinfo {author} {\bibfnamefont {L.~R.}\ \bibnamefont
  {Lawlor}}\ and\ \bibinfo {author} {\bibfnamefont {J.~M.}\ \bibnamefont
  {Smith}},\ }\bibfield  {title} {\bibinfo {title} {The coevolution and
  stability of competing species},\ }\href {https://doi.org/10.1086/283049}
  {\bibfield  {journal} {\bibinfo  {journal} {The American Naturalist}\
  }\textbf {\bibinfo {volume} {110}},\ \bibinfo {pages} {79–99} (\bibinfo
  {year} {1976})}\BibitemShut {NoStop}%
\bibitem [{\citenamefont {Grunert}\ \emph {et~al.}(2021)\citenamefont
  {Grunert}, \citenamefont {Holden}, \citenamefont {Jakobsen},\ and\
  \citenamefont {Stenseth}}]{Grunert2021}%
  \BibitemOpen
  \bibfield  {author} {\bibinfo {author} {\bibfnamefont {K.}~\bibnamefont
  {Grunert}}, \bibinfo {author} {\bibfnamefont {H.}~\bibnamefont {Holden}},
  \bibinfo {author} {\bibfnamefont {E.~R.}\ \bibnamefont {Jakobsen}},\ and\
  \bibinfo {author} {\bibfnamefont {N.~C.}\ \bibnamefont {Stenseth}},\
  }\bibfield  {title} {\bibinfo {title} {Evolutionarily stable strategies in
  stable and periodically fluctuating populations: The rosenzweig–macarthur
  predator–prey model},\ }\bibfield  {journal} {\bibinfo  {journal}
  {Proceedings of the National Academy of Sciences}\ }\textbf {\bibinfo
  {volume} {118}},\ \href {https://doi.org/10.1073/pnas.2017463118}
  {10.1073/pnas.2017463118} (\bibinfo {year} {2021})\BibitemShut {NoStop}%
\bibitem [{\citenamefont {Bishop}\ and\ \citenamefont
  {Cannings}(1978)}]{Bishop1978}%
  \BibitemOpen
  \bibfield  {author} {\bibinfo {author} {\bibfnamefont {D.~T.}\ \bibnamefont
  {Bishop}}\ and\ \bibinfo {author} {\bibfnamefont {C.}~\bibnamefont
  {Cannings}},\ }\bibfield  {title} {\bibinfo {title} {A generalized war of
  attrition},\ }\href {https://doi.org/10.1016/0022-5193(78)90304-1} {\bibfield
   {journal} {\bibinfo  {journal} {Journal of Theoretical Biology}\ }\textbf
  {\bibinfo {volume} {70}},\ \bibinfo {pages} {85} (\bibinfo {year}
  {1978})}\BibitemShut {NoStop}%
\bibitem [{\citenamefont {Wiggins}(2003)}]{wiggins_2003}%
  \BibitemOpen
  \bibfield  {author} {\bibinfo {author} {\bibfnamefont {S.}~\bibnamefont
  {Wiggins}},\ }\href@noop {} {\emph {\bibinfo {title} {Introduction to Applied
  Nonlinear Dynamical Systems and Chaos}}}\ (\bibinfo  {publisher}
  {Springer-Verlag},\ \bibinfo {year} {2003})\BibitemShut {NoStop}%
\end{thebibliography}%

\end{document}